\documentstyle[12pt,psfig,epsfig]{article}
\begin{document}     

\title{Hadronic Interactions and TeV Muons
\\ in Cosmic Ray Cascades}
\author{G.  Battistoni} 
\date{}

\maketitle

\begin{center}

{\small {\it I.N.F.N., Sezione di Milano, via Celoria 16, I-20133 Milano, Italy}} \\

\end{center}

\begin{abstract}
In view of the interpretation of data collected by large deep underground
experiments in terms of primary Cosmic Ray physics, 
this work is focused on the study of the production of  
of TeV muons in Extensive Air Showers.
The review tries to point out those features of hadronic interactions that 
mostly affect the production of the high energy muons.
A few different Monte Carlo codes are compared, with a particular attention
to those based on the Regge-Gribov framework.
The possibility of performing experimental tests 
of the proposed models is also briefly discussed.
\end{abstract}


\section{Introduction}

Cosmic rays of the highest energies can be observed only through
their interaction with the  Earth's atmosphere: the reliability of our
interpretation of the features of such secondary  particles, and of their
relation to the characteristics of the primary particle, is necessarily 
related to quality of our understanding of hadron--hadron, hadron--nucleus and
nucleus--nucleus interactions. This aspect is particularly stimulating for high
energy physicists,  since there is not yet an exact way to calculate the
properties of the bulk of hadronic  interactions. Also, from the experimental
point of view, the productions of secondary  cosmic rays at very high energies
occurs in kinematic regions, or energy ranges, that have  not been explored
in accelerator experiments, and that will hardly be accessed 
even at the  hadron
colliders of the next generation. Therefore, Extensive Air Showers
 of energy $>10^{15}$
still represent an almost unique chance to test our theoretical achievements
in very high energy nuclear physics, not to speak of the region above 
$10^{17}$ eV, which is, at least now, completely out of reach of the present
accelerator technology.

In the last years, new experiments, located on the earth surface or underground,
have collected valuable data concerning different Extensive Air Shower (EAS) components. 
The question of the interpretation of these
data, the choice of the best possible simulation tool, and in general
the evaluation of the systematics associated to such simulations,
has become crucial. This remains a fundamental point also for the 
future activities at the extreme high energies\cite{auger}.

A decisive contribution to this discussion has come from the work
carried on in the Karlsruhe group\cite{kaskade}, concerning
the maintenance and distribution of the CORSIKA code\cite{corsika},
which now can be interfaced to different hadronic packages.
A detailed comparison of the results from these models
has been presented in \cite{karls}, and new progresses
have been also presented at this conference\cite{knapp}.
There, the emphasis was given to the data collected
by surface arrays, mostly the e.m. component and
the low energy (Gev) muons.

The purpose of this work is to present a further
complement to the discussion started in ref.\cite{karls}, 
focusing the review on the Monte Carlo predictions
which are specific for the production of very high energy muons (in the
TeV range), as those collected by the large underground
experiments.
This is topic is of particular relevance in the indirect analysis of
EAS, since TeV muons come from the decay of mesons
produced in the very early stages of the cascade,
and are more directly related to the properties of high energy interaction
with respect to the low energy muon component.

After the pioneristic work of Utah's experiment\cite{utah}
and the results from NUSEX\cite{nusex} and FREJUS\cite{frejus},
recent experimental papers on the subject have been published by
experiments at Gran Sasso, like MACRO\cite{macrocomp,macrocomp2}, and also by 
Soudan2\cite{soudan2} and KGF\cite{kgf}. The MACRO data are important
for the large collection area and the statistics, apart from the interest in
the analysis method proposed in \cite{macrocomp2}.  
The relevance of KGF results is mainly the high energy selection because of the
large rock overburden. Both experiment based their analyses on 
Monte Carlo models based on parameterizations experimental results 
at colliders. A critical discussion of these models is now considered
mandatory.
Some of these experiments, like MACRO or LVD\cite{lvd}, 
take also data in coincidence\cite{coinc} with the 
EAS-TOP surface array\cite{eastop}.
For this reason, the calculation of the TeV muon component has to 
be studied also in correlation with other EAS components.
This will be the subject of a much more complete review\cite{aabfr}, of
which the present work is just a partial anticipation.

In the next section the features of hadronic interactions which are
relevant for muon production are reviewed. Then, in Section 3,
the Monte Carlo generators under study are briefly
described. Some results about the simulation of primary interactions are
given in Section 4, while the comparison of full shower calculations
are presented in Section 5. A general discussion is attempted in
the Conclusions.

\section{High energy muons as a probe of hadronic interactions}

The high energy muons in EAS come from the decay of pions and kaons produced
in the early stages of shower development. 
The interest is therefore in the cross sections, multiplicities and their
fluctuations, for meson production in all kinds of possible 
hadron and nucleus interactions in the relevant energy range.
A key aspect is the competition of decay and interaction of these mesons along
their path in a medium with varying density profile, such as 
the atmosphere.
Therefore, while the inelastic cross section of primaries determines
the height of first interaction, the inelastic cross sections
of mesons determine the interaction length. The decay length is
a function of meson energy, whose distribution is governed
by the $X_{F}$ distribution
as resulting
from meson, nucleon or nucleus collisions with air nuclei.
In particular, the high $X_{F}$ values (``fragmentation region''), 
poorly studied in accelerator experiments, are the dominant ones.
Diffractive and non diffractive interactions are both to be considered.

The $P_\perp$ distribution affects the transverse structure of muons
in the showers, since the kink in the meson decay can be considered
negligible. At TeV energies, geomagnetic deflection
is a minor correction. A realistic calculation must take into account
the correlations existing between $P_\perp$, energy and rapidity (or
$X_{F}$). As already discussed in \cite{vulc_gais}, while a 
simultaneous change in the interaction lengths of mesons and nucleons
produces only second order effects in the eventual muon yield, 
variations in $P_\perp$ are directly reflected in the calculation
of the lateral distribution. 

Of course, the lateral distribution is always
the result of a convolution between longitudinal and transverse
properties of the interaction. Calling $r$ the displacement
of a muon with respect to the shower axis, as measured at a 
slant distance $H_{prod}$ from the relevant interaction, then:
\begin{equation}
r \sim \frac{P_\perp}{E_{\pi,K}} H_{prod}
\end{equation}
In this simplified description the transverse momentum in the parent decay
is neglected.
The previous expression can be re--expressed in a more instructive
way, considering that at high energy 
the longitudinal c.m. variable $X_F$ is approximately
equal to the laboratory energy fraction
$X_{lab}$ in the forward region, up
to terms of the order of $(m_\perp/E_0)^2$) :
\begin{eqnarray}
& r \sim  \frac{P_\perp}{X_F^{\pi,K}E_0} H_{prod}  \nonumber \\
& \propto \frac{P_\perp}{X_F^{\pi,K}E_0} 
\left( \log{\sigma^{inel}_{n-Air}}+const.\right) 
\end{eqnarray}
where a simple exponential atmosphere profile
has been used in the last expression. 

Of course, in the case of underground experiments, muon propagation 
in the rock is also a fundamental aspect of calculation, but it
is a factorizable ingredient, not to be discussed in this work.
A discussion of this topic can be found in \cite{muontransp}.
It can just be mentioned that, even for large depths, although
the scattering in the rock is quantitatively important, the
simulation results show that the amount of this deviations it is not yet
enough to obscure the influence of $P_\perp$ in the
lateral distribution of the surviving muon component.

\begin{figure}[p]
\begin{center}
\mbox{\epsfig{file=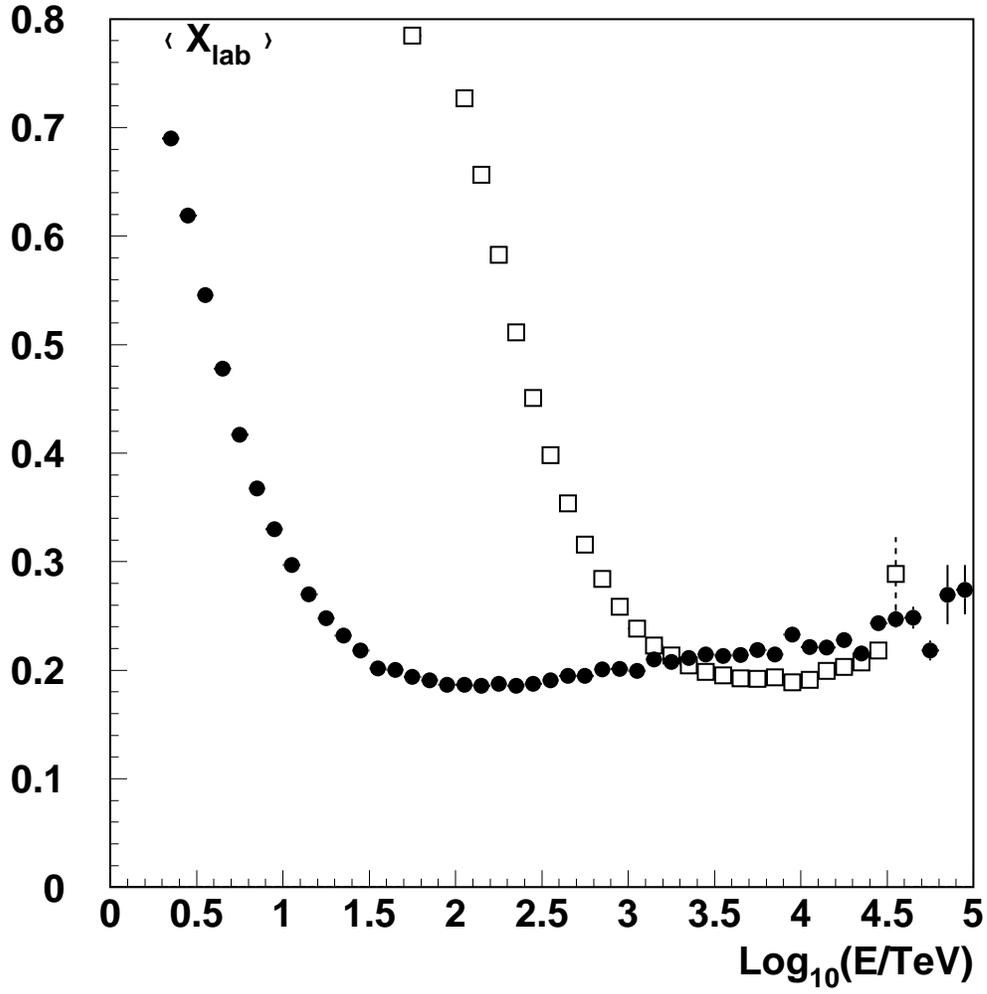,width=15.0cm}}
\caption{Monte Carlo calculation of
Average $X_{lab}$ of parent pions of muons detected
after 3400 hg/cm$^2$ of standard rock, at 30$^\circ$ of zenith angle,
for proton (black circles) and iron nuclei (open squares) primaries.
\label{fi:f1}}
\end{center}
\end{figure}

The energy filter provided by the rock, imposes a threshold on the muon energy,
and therefore also on its parents' energy. This introduces some selection 
effects. Considering for example the case of Gran Sasso laboratory.
There, the minimum rock overburden is about 3100 hg/cm$^2$ (in the direction
of Campo Imperatore, where EAS--TOP is located). That depth corresponds to
an energy threshold for muons of about $E_\mu^{thr}~\sim$ 1.3 TeV. 
For an integral spectral index of primaries $\gamma$=1.7, 
the corresponding average energy of parent pions is
\begin{equation}
<E_\pi> = \frac{\gamma+2}{\gamma+1}E_\mu^{thr} 
\frac{1-{r_\pi}^{(\gamma+1)}}
{1 - {r_\pi}^{(\gamma+2)}} 
\simeq 1.6~TeV
\end{equation}
where $r_\pi$=$m_\mu^2/m_\pi^2$.
When considering the first primary interaction of a nucleon, at energy $E_0$,
a threshold in muon energy is thus equivalent to a threshold in 
$X_{F} \simeq X_{lab} = \frac{E_\pi}{E_0}$, 
which for ``low'' $E_0$ values (a few TeV) can be close to 1, {\it i.e.}
in the very forward region.
In that cases, in practice only muons from the first generation in the cascade
are detected, and the primary interaction is really investigated.
At fixed threshold, for increasing primary energy, the average $X_{lab}$
of parents at first decreases, and then exhibits a smooth rise, when 
more and more cascade generation can provide mesons above threshold.
This is shown in Fig.\ref{fi:f1} for parent pions from
 proton and Fe primaries, in the 
case of 3400 hg/cm$^2$, at 30$^\circ$ of zenith angle.

Since below 10 TeV, the composition of primaries contributing
to TeV muons is strongly dominated by protons, this selection effect
is being exploited in the coincidence experiment of EAS--TOP
and MACRO to investigate the differences between interaction models
in the fragmentation region\cite{coinc2}.

In this limited discussion the interesting question of
prompt muons from the decay of heavy flavored mesons will be neglected. 
However, it is recognized that this production  is indeed a very small 
fraction of TeV muons surviving underground 
(see for instance ref. \cite{charm}), and there is debate on the real
possibility of a positive detection of this component.

\section{Monte Carlo generators}

After the first fundamental work of Elbert\cite{elbert},
an important paper concerning the simulation of muons
detected deep underground is that of ref. \cite{nim85}, where 
parameterizations as a function
of energy per nucleon, zenith angle and rock depth were presented,
following the method already pointed out in\cite{elbert}.
The calculations were performed using, as interaction model,the
``splitting Hillas' algorithm''\cite{hillas}. 
This algorithm is indeed simple, perfectly preserving Feynman scaling
and does not provide any information about the transverse structure
of the showers. 
More realistic predictions were obtained with the development of the
HEMAS code\cite{hemas}, which was explicitly designed to provide
a dedicated tool for the high energy muon component (above 500 GeV), in view
of the experimental activity at Gran Sasso laboratory.
The hadronic interaction model of HEMAS is based on the 
parameterizations of collider data, mostly UA5\cite{UA5}, following a
picture in which clusters of particles were
formed in the nucleon--nucleon collisions, 
eventually decaying into ordinary mesons.
Effects due to the nuclear nature of the target, as measured in heavy ion 
physics, are used to correct multiplicity, transverse momentum, etc.

The most noticeable differences coming from these collider 
parameterizations concern two aspects:
\begin{enumerate}
\item the larger fluctuations of muon multiplicity, in connection
to the negative binomial multiplicity distribution of produced
mesons
\item the transverse distribution of muons, following 
the power law distribution related to that of
$P_\perp$ and its correlation with 
energy and multiplicity.
\end{enumerate}

The first analyses by MACRO\cite{macrocomp,macrodeco1} 
could establish the much better
description of data provided by HEMAS with respect to the 
previous parameterizations of \cite{nim85}. 
It must be quoted that also the recent analysis of KGF made use
of similar collider parameterizations.

In any case, this type of codes have still a weak point.
The application to cosmic rays necessarily requires
some extrapolation, in particular towards the high $X_F$ values, and this
procedure has always some degree of arbitrariness. 
Also, in the construction of parameterizations, important correlations
among variables might be lost. 
These are some of the reason that have driven people to look for
``theoretically inspired'' models, with a limited number of parameters
and a phenomenological framework as a guidance.
A rather remarkable success in hadronic physics seems to be 
achieved by the Dual Parton Model\cite{dpm} (DPM).
This approach makes use of concepts
derived from the mathematical requirements
of scattering theory, as unitarity and analyticity 
together
with color flow and parton idea and topological expansion of QCD.
Such concepts can be formulated in the framework of
relativist quantum field theory, according to the
Regge--Gribov approach\cite{grt}.
The extension from nucleon--nucleon to nucleon--nucleus
and nucleus--nucleus is achieved according to the
Glauber multiple scattering formalism\cite{glauber}.
A last fundamental ingredient of all these models is the ``hadronization''
algorithm,  where parton strings produced in the scattering
process are converted into ordinary particles.
Some authors make use of the hadronization Lund models\cite{lund}.

In practice, many different variations exist of these
basic ideas, and also many different practical implementation
have been developed in different numerical codes. 
Some of these have been successfully interfaced
to the general purpose shower code CORSIKA, as already 
mentioned in the introduction.
A detailed description of these models can be found only in the original 
literature, however the most important
features are the following\footnote{CORSIKA makes use of the 
 GHEISHA\cite{gheisha} parameterized model below 80 GeV incident 
hadron energy.
This is irrelevant for TeV muons.}.
\begin{enumerate}
\item HDPM\cite{corsika}. This is the original interaction package
contained in CORSIKA, where parameterizations of the main results
of DPM are contained. It will not be considered in the following.

\item SIBYLL\cite{sibyll}. It was explicitly developed for cosmic
ray application. It contains DPM ideas, but the present version is
deeply based on the minijet production as calculable from QCD, providing
all the rising part as a function of energy of the nucleon--nucleon
cross section. The soft part is constant. A Lund code for
hadronization is used.

\item DPMJET\cite{sibyll}. It is based on the ``Two component''
DPM, {\i.e.} both soft processes and hard QCD scattering
are unitarized together. The soft part assumes the ``supercritical
pomeron'' so that the soft contribution of the cross section is not constant
but rises as $s^{0.08}$. Multi-Pomeron exchange is possible.
A specific point of merit of this code is the inclusion
of dedicated algorithms to treat nuclear effects.
Also DPMJET makes use of a Lund hadronization code.
It must be mentioned that 
this interaction model has been also chosen for a new 
version of HEMAS\cite{hemasdpm}, called HEMAS--DPM,
again specifically devoted to high energy muons. This is presently
in use by the MACRO experiment, and preliminary results are
presented at this conference\cite{ornella}.
From the point of view of results, no significant
differences with respect to those achieved inside CORSIKA
are expected, although
the comparison of results from the same interaction model from
two different shower codes would be a technically interesting
issue. This will be discussed in \cite{aabfr}.

\item QGSJET\cite{qgsjet}. It is based on the
``Quark Gluon String'' model\cite{qgs}.
It is very similar to DPMJET as far the
pomeron parameters are concerned. It also includes the contribution
of perturbative QCD. At present there are important differences
with respect to DPMJET in the way in which soft and hard scattering
are merged together.

\item VENUS\cite{venus}. Originally developed for accelerator
physics. It contains DPM with the supercritical pomeron, and the
addition of some specific features concerning color exchange processes.
It introduces secondary interaction of strings before hadronization.
The version at present in CORSIKA does not yet 
include the contribution of minijets.

\end{enumerate}

In the following results concerning high energy muons
from some of these choices will be compared. 
For some specific issues the with the original HEMAS
code will be also shown.

\section{Particle Production in p--Air single interaction}

The first approach in the comparison of models for shower calculation
is the analysis of the features of primary interaction.
In principle,  one has to study the primary 
interaction in both nucleon--Air and nucleus--Air collisions.
However, for simplicity, the nucleon--Air interaction can be considered
as the basic representative ingredient of the models. 
The extension to
nucleus--nucleus interaction according to
the Glauber model could be, in principle, a
common part to all codes. 
However,
a check of how this is realized in practice
 is mandatory and will be addressed in future.
For the comprehension of shower development, 
the study of meson--Air interaction would be also important.
An exhaustive examination would require a ponderous dissertation,
that goes beyond the present possibility.
The discussion will be limited 
to the inelastic cross section, the $X_{lab}$ and
the $P_\perp$ distributions.

\subsection{The inelastic cross section}

\begin{figure}[p]
\begin{center}
\epsfig{file=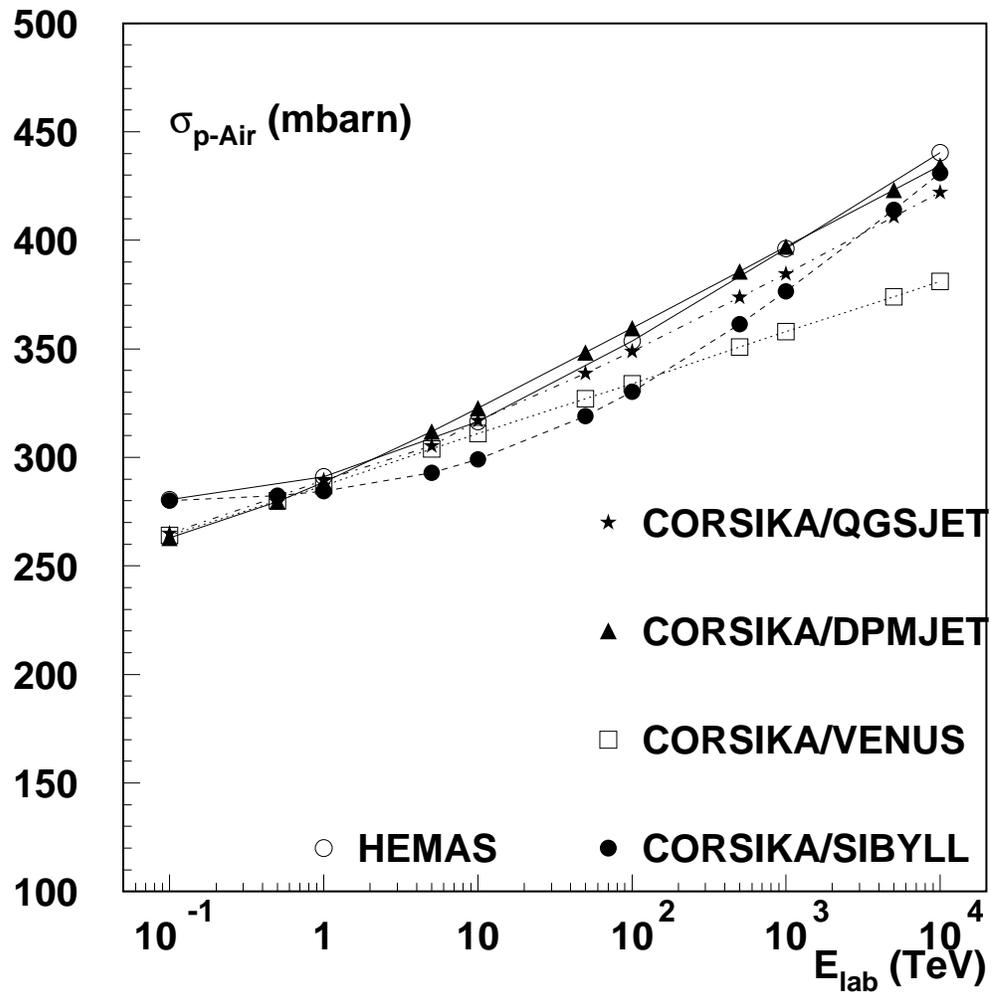,width=15.0cm}
\caption{
Inelastic p--Air cross section as a function of laboratory
energy for some of the considered Monte Carlo models.
\label{fi:f2}}
\end{center}
\end{figure}

In Fig. \ref{fi:f2}, the p--Air inelastic cross section as a function
of laboratory energy is shown for some of the considered models, in the
relevant energy range accessed with muons detected deep
underground.

There are significative differences in the knee region (VENUS seems to
depart more from the other ones), but also at 
$\sqrt s$ values where good p--p data are available.
Here the SIBYLL cross section seems to exhibit the major deviation.
Presumably, this can be attributed to a non correct inclusion of
diffractive events in the Glauber calculation\cite{stanev}, and in the
next version of SIBYLL this problem will be corrected.
These differences in cross section will be reflected by different
average interaction heights and some change in the muon production
height.

\subsection{Inclusive distributions and the Spectrum--weighted
moments}

\begin{figure}[p]
\begin{center}
\mbox{\epsfig{file=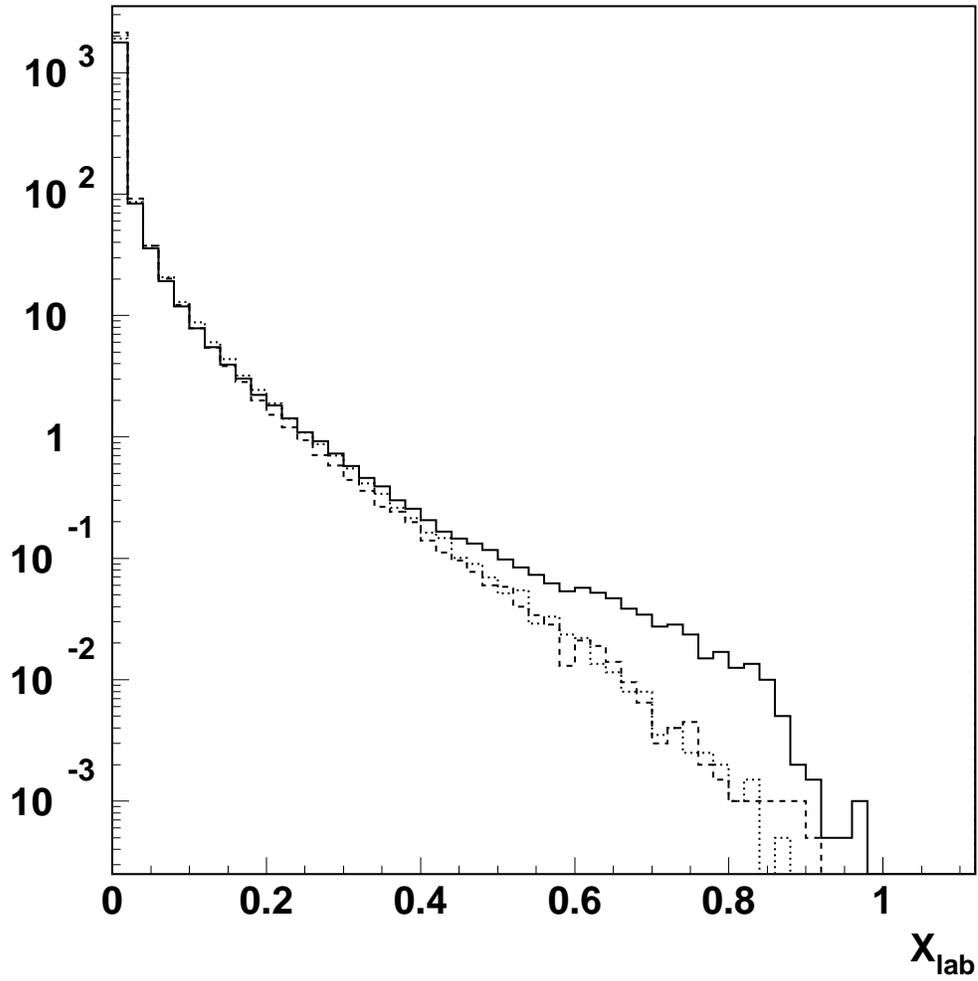,width=15.0cm}}
\caption{
$X_{lab}$ distribution for charged pions produced
in p--Air collisions at 200 TeV in the laboratory
frame for DPMJET (continuous line), QGSJET (dashed) and VENUS (dotted).\label{fi:fi3}}
\end{center}
\end{figure}

As a significant example of possible difference between codes, 
Fig. \ref{fi:fi3} shows the $X_{lab}$ 
distribution of charged pions produced
in p--Air collisions at 200 TeV in the laboratory frame, for 3 different
codes: DPMJET (continuous line), QGSJET (dashed) and VENUS (dotted).
The difference of DPMJET with respect the other codes at large
$X$ is due to a particular mechanism of Lund fragmentation (JETSET 7.3) of the
fast valence ``diquark''\footnote{{\it i.e.} the projectile after the stripping of
one valence quark} in the projectile, in the case of positive pions (``popcorn''
effect\cite{popcorn}). No effect of this kind is instead possible for kaon production, where
strange quark are not presented in the valence quarks, or diquarks, of
the incoming nucleon.
This mechanism is not included in the other
MC codes.
This feature has indeed relevance in the yield of high energy muon.
It is important to notice that, apart from the region near $X_{lab}$=0
and $X_{lab}$=1, these Regge--Gribov model exhibit a substantial Feynman scaling.
Therefore these features are preserved in a wide range of energies.

\begin{figure}[p]
\begin{center}
\mbox{\epsfig{file=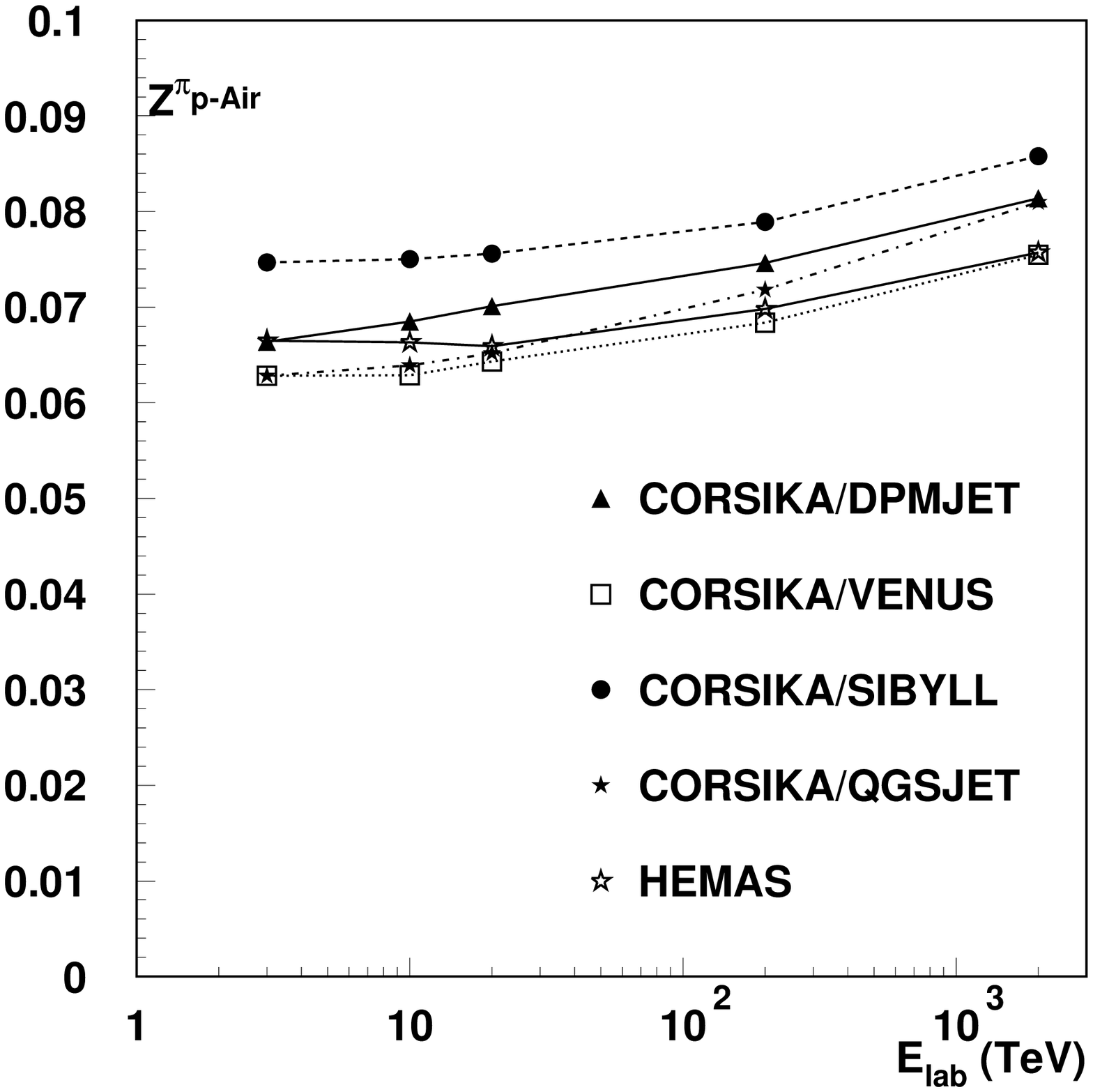,width=15.0cm}}
\caption{
Comparison of Z moments for charged pion
production as a function of energy,
for different codes.\label{fi:fi4}}
\end{center}
\end{figure}

In the current literature, some emphasis has been given to the
``spectrum--weighted'' moments, defined, for instance in case
of pion production, as
\begin{equation}
Z^{\pi} = \int_{0}^{1} (X_{lab})^{\gamma - 1} \frac{dN^{p+Air\rightarrow \pi + X}}{dX_{lab}} dX_{lab}
\end{equation}
where $\gamma$ is spectral index of primary nucleons. Such a function give a measurement
of the inclusive particle yield in cosmic ray showers after the integration over the energy
spectrum\cite{zmom}. 
In principle, the high energy muon yield is proportional to those values.
However, the comparison of Z-moment from different codes, as that shown in Fig. \ref{fi:fi4},
cannot give completely meaningful information when the inelastic cross sections are different.
In fact, a larger Z factor (as in the case of SIBYLL) will not
give a larger muon yield if the production height is lower,
so that available length for decay is smaller. 

\begin{figure}[p]
\begin{center}
\mbox{\epsfig{file=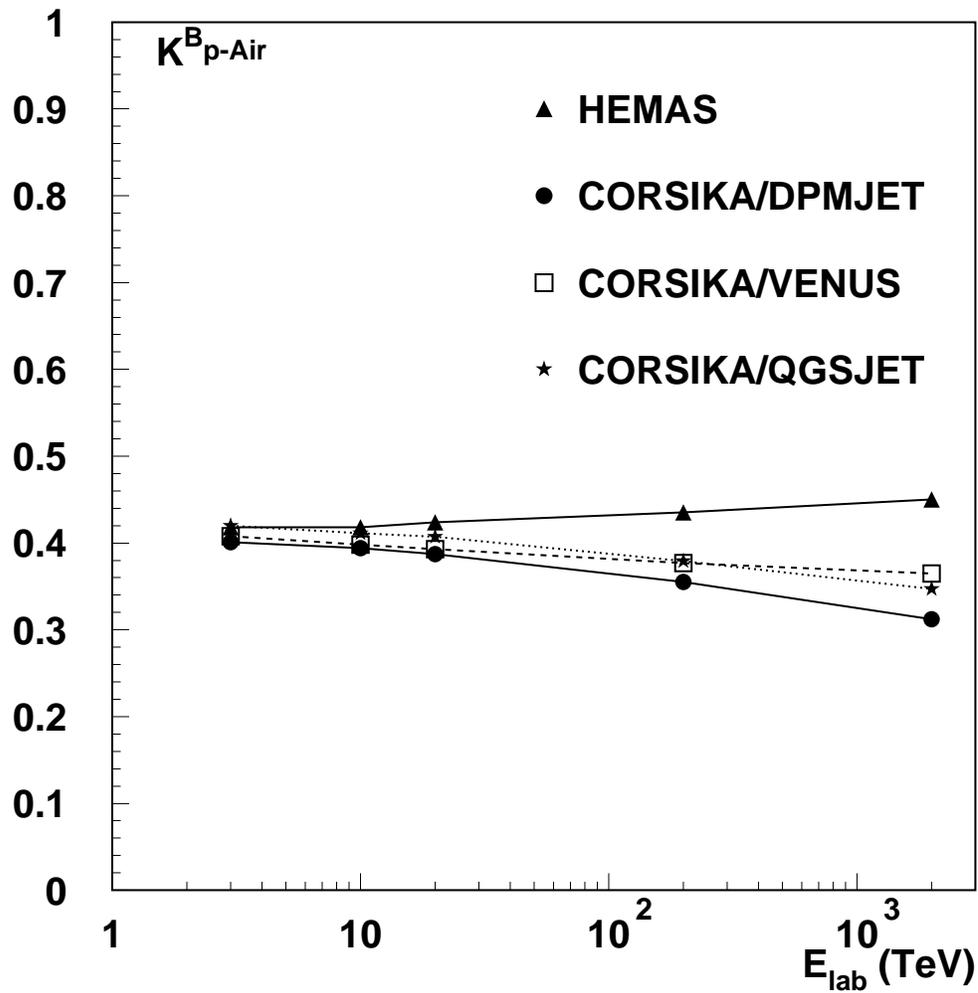,width=15.0cm}}
\caption{
Comparison of the energy fraction carried away
by leading nucleons 
in p--Air collisions
as a function of laboratory energy, for
different models. \label{fi:fi5bis}}
\end{center}
\end{figure}

It is also interesting to inspect the behaviour of the 
energy fraction $K_B$ carried away by the leading 
baryon in p--Air collisions as a function of primary energy.
This is shown in Fig.\ref{fi:fi5bis}.
It can be seen how the HEMAS code exhibits a different
behaviour 
from other representative models of the QGS and 
DPM kind, where elasticity smoothly decreases with energy.
In this respect, it is worthwhile to remark what stated 
in ref.\cite{gaisser97} about the cluster models.
Such approaches have the general tendency to produce events
which become increasingly elastic as energy increase\footnote{It has been suggested
that a less unbiased way to look at elasticity would be to consider
$K_{B-\bar B}$, where anti--baryons are taken with the minus sign.}.
In the same reference it was shown how QGSJET 
and DPM model represent interactions on nuclear 
targets in a way which is consistent with low 
energy data.
It can be important to notice how a model
like the quoted Hillas' algorithm, in its original form, would give a
constant $K_B$=0.5.

\begin{figure}[p]
\begin{center}
\mbox{\epsfig{file=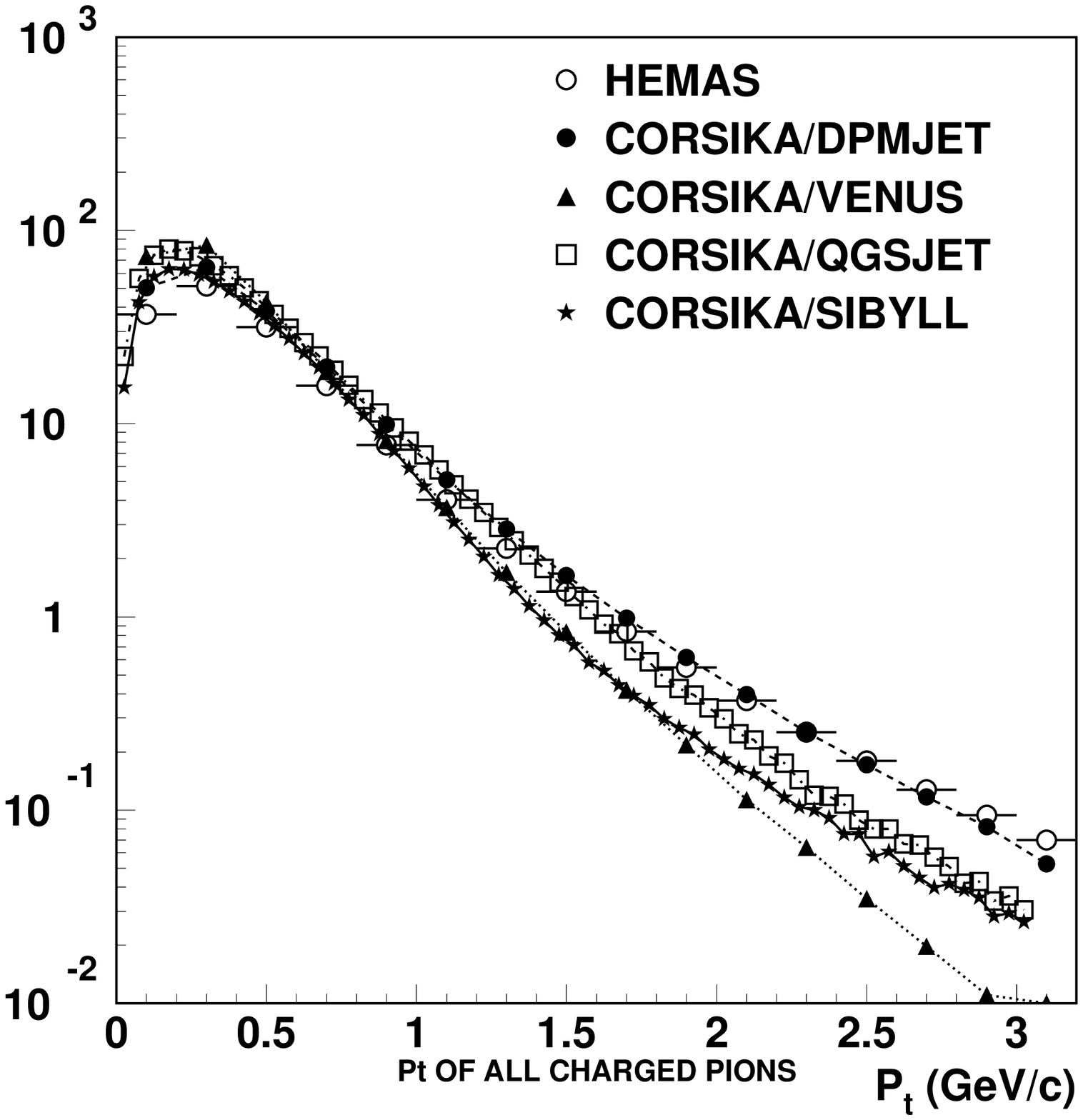,width=15.0cm}}
\caption{
Comparison of $P_\perp$ distributions of
charged pions produced in p--Air collisions
at 200 TeV laboratory energy, for
different models. \label{fi:fi5}}
\end{center}
\end{figure}

A few remarks are necessary as far as the $P_\perp$ distribution
is concerned.
The phenomenological framework of the
Regge--Gribov theories, by its nature,
provides only predictions for the longitudinal properties of the interaction.
The transverse structure leading to the specific $P_\perp$ distribution is not
constrained by the theory, but for the higher $P_\perp$ phenomena, where
perturbative QCD can be used (of small relevance in the primary energy region
addressed in general by the underground experiment).
Once again, the model builders have to be guided mostly by experimental data,
introducing {\it a-priori} determined functional forms with their additional
parameters. 
Again, as a typical example, 
the $P_\perp$ distribution
of charged pions produced in p--Air collisions
at 200 TeV laboratory energy, as obtained in 
different models is shown 
in Fig.\ref{fi:fi5}.

It can be pointed out how all the considered
models produce $P_\perp$ distributions which, following the
experimental results at colliders, can be fitted
with a power law spectrum.
It can be also noticed how HEMAS and DPMJET 
give very close results, predicting larger tails than the other 
models at relatively high values of $P_\perp$.

\section{Shower Calculations}

Test simulation runs of
EAS development have been performed, choosing typical parameters for
the experimental situation at Gran Sasso laboratory.
According to the logic followed so far, 
results are presented here only for the case of proton primaries.
Runs have been performed 
Runs were performed at different fixed energies,
for fixed angles (30$^\circ$ zenith and
190$^\circ$ azimuth),
setting the observation level for EAS at 2000 m a.s.l., and
with the geomagnetic field specific for the site.
The chosen direction corresponds to the typical one 
from which the underground halls see the EAS--TOP array.
Propagation of TeV muons in rock has been calculated by means
of the PROPMU code\cite{lipstan} for the corresponding thickness
of 3400 hg/cm$^2$ of standard rock. 
In Table \ref{tb:t1} the
average yield of muons as measured at the surface (E$_\mu\geq$1 TeV)
is reported, together with 
that of the survived muons underground.

\begin{table} [p]
\newlength{\digitwidth} \settowidth{\digitwidth}{\rm 0}
\catcode`?=\active \def?{\kern\digitwidth}
\caption{Comparison of average TeV muon yield at 2000 m a.s.l.
and of the no. of muons surviving underground, after 3400 hg/cm$^2$,
produced in proton induced showers at different energies.
The same muon transport code has been applied in all runs.
The relative statistical error on the reported figures 
is of the order of 5
\label{tb:t1}}

\begin{tabular*}{\textwidth}{@{}l@{\extracolsep{\fill}}cc}
\hline
                  \multicolumn{3}{c}{p--Air, 3 TeV} \\
\hline
Code    & $<N_\mu(E\geq1~TeV)>$   & $<N_\mu>$ survived undergr. \\
\hline
HEMAS           &  0.0027   &  0.00049                      \\
CORSIKA/DPMJET  &  0.0035  &  0.00059                       \\
CORSIKA/QGSJET  &  0.0018   &  0.00015                     \\
CORSIKA/VENUS   &  0.0028   &  0.00031                     \\
CORSIKA/SIBYLL  &  0.0042   &  0.00056                  \\
\hline
\end{tabular*}
\begin{tabular*}{\textwidth}{@{}l@{\extracolsep{\fill}}cc}
\hline
                  \multicolumn{3}{c}{p--Air, 10 TeV} \\
\hline
Code    & $<N_\mu(E\geq1~TeV)>$   & $<N_\mu>$ survived undergr. \\
\hline
HEMAS           &  0.064   &  0.015 \\
CORSIKA/DPMJET  & 0.055  &  0.016 \\
CORSIKA/QGSJET  &  0.056 &  0.015 \\
CORSIKA/VENUS   &  0.057 &  0.015 \\
CORSIKA/SIBYLL  &  0.064   &  0.018 \\
\hline
\end{tabular*}
\begin{tabular*}{\textwidth}{@{}l@{\extracolsep{\fill}}cc}
\hline
                  \multicolumn{3}{c}{p--Air, 20 TeV} \\
\hline
Code    & $<N_\mu(E\geq1~TeV)>$   & $<N_\mu>$ survived undergr. \\
\hline
HEMAS           &  0.182 &  0.048  \\
CORSIKA/DPMJET  &  0.142   &  0.049  \\
CORSIKA/QGSJET  &  0.153 &  0.052  \\
CORSIKA/VENUS   &  0.150  &  0.049  \\
CORSIKA/SIBYLL  &  0.153   &  0.053\\
\hline
\end{tabular*}
\begin{tabular*}{\textwidth}{@{}l@{\extracolsep{\fill}}cc}
\hline
                  \multicolumn{3}{c}{p--Air, 200 TeV} \\
\hline
Code    & $<N_\mu(E\geq1~TeV)>$   & $<N_\mu>$ survived undergr. \\
\hline
HEMAS           &  1.29  &  0.47  \\
CORSIKA/DPMJET  &  1.17   &  0.49  \\
CORSIKA/QGSJET  &  1.17   &  0.48  \\
CORSIKA/VENUS   &  1.28   &  0.52  \\
CORSIKA/SIBYLL  &  1.11   &  0.44  \\
\hline
\end{tabular*}
\end{table}

\begin{table}[p]
\catcode`?=\active \def?{\kern\digitwidth}
\caption{Comparison of average TeV muon yield at 2000 m a.s.l.
and of the no. of muons surviving underground, after 3400 hg/cm$^2$,
produced in showers induced by different primary nuclei at 2000 TeV total energy.
\label{tb:t2}}
\begin{tabular*}{\textwidth}{@{}l@{\extracolsep{\fill}}cc}
\hline
                  \multicolumn{3}{c}{p--Air, 2000 TeV} \\
\hline
Code    & $<N_\mu(E\geq1~TeV)>$  &   $<N_\mu>$ survived undergr. \\
\hline
HEMAS           &  6.44   &  2.55  \\
CORSIKA/DPMJET  &  6.90   &  2.84  \\
CORSIKA/QGSJET  &  6.79   &  2.80  \\
CORSIKA/VENUS   &  7.43   &  3.09  \\
CORSIKA/SIBYLL  &  5.99   &  2.48  \\
\hline
\end{tabular*}
\begin{tabular*}{\textwidth}{@{}l@{\extracolsep{\fill}}cc}
\hline
                  \multicolumn{3}{c}{He--Air, 2000 TeV} \\
\hline
Code    & $<N_\mu(E\geq1~TeV)>$   & $<N_\mu>$ survived undergr. \\
\hline
HEMAS           &  9.51   &  3.75  \\
CORSIKA/DPMJET  &  9.61   &  3.94  \\
CORSIKA/QGSJET  &  8.89   &  3.66  \\
CORSIKA/VENUS   &  10.26   & 4.23  \\
CORSIKA/SIBYLL  &  8.67   &  3.59  \\
\hline
\end{tabular*}
\begin{tabular*}{\textwidth}{@{}l@{\extracolsep{\fill}}cc}
\hline
                  \multicolumn{3}{c}{Fe--Air, 2000 TeV} \\
\hline
Code    & $<N_\mu(E\geq1~TeV)>$  & $<N_\mu>$ survived undergr. \\
\hline
HEMAS           &  16.42   &  5.61 \\
CORSIKA/DPMJET  &  15.72   &  5.86  \\
CORSIKA/QGSJET  &  15.87   &  5.74  \\
CORSIKA/VENUS   &  16.92   &  6.17  \\
CORSIKA/SIBYLL  &  15.15   &  5.57  \\
\hline
\end{tabular*}
\end{table}

\begin{figure}[p]
\begin{center}
\mbox{\epsfig{file=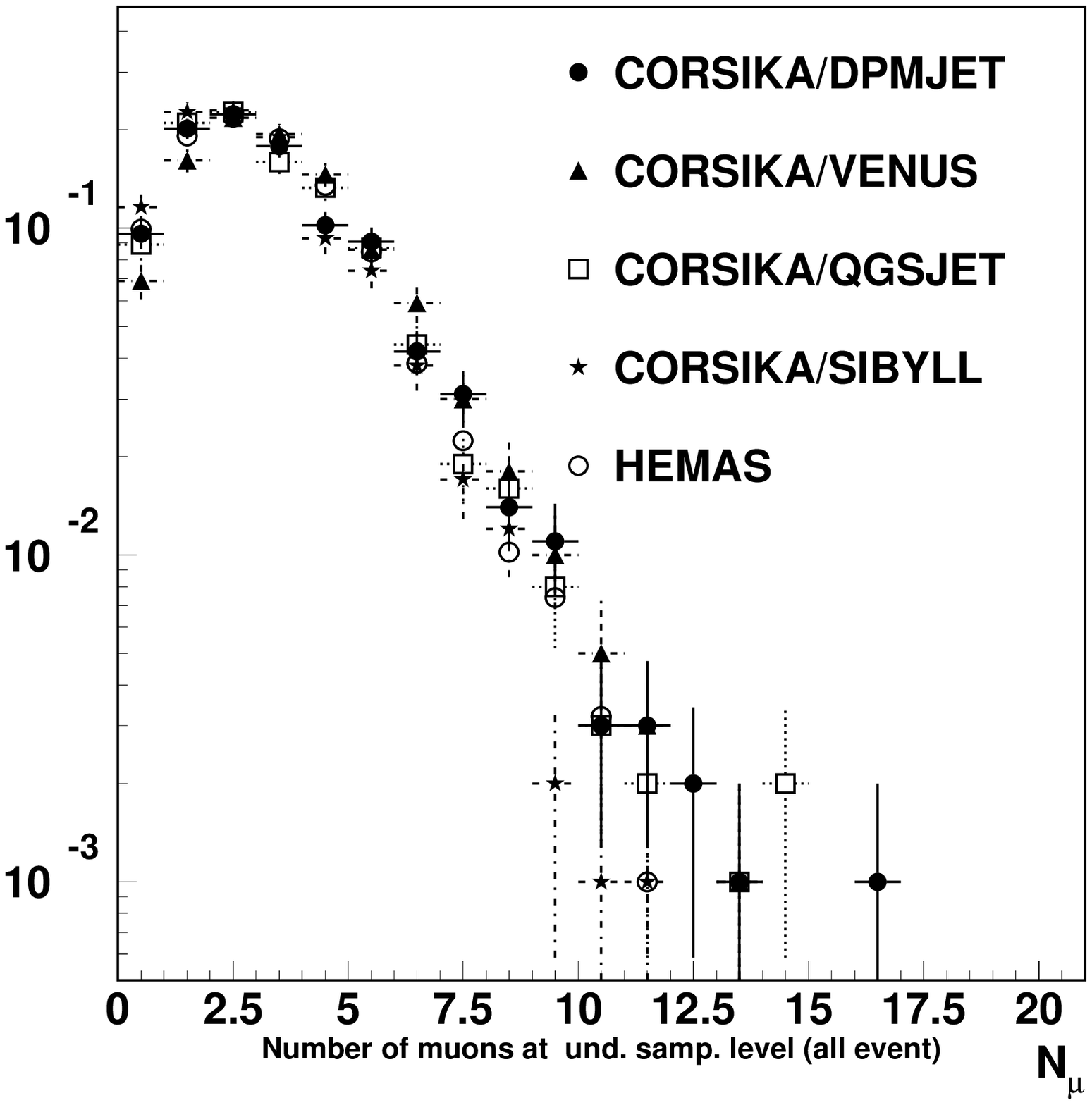,width=15.0cm}}
\caption{
Comparison of normalized distributions 
of underground muon multiplicity 
from p--Air collisions
at 2000 TeV laboratory energy, 30$\circ$
zenith angle, and 3400 hg/cm$^2$ depth, for
different models. \label{fi:mu}}
\end{center}
\end{figure}

A comparison of muon yields in showers from protons and primary nuclei
is shown, for example at 2000 TeV total energy, in Table \ref{tb:t2}.
The shape of the muon multiplicity distributions from the
different models are very similar, as, for instance, in the
example of Fig. \ref{fi:mu}, from p--Air collision at 2000 TeV.

As far as the underground muon yield is concerned,
it can be seen how these predictions are in general close
one to the other. The most relevant percentage differences, for a given threshold
on muon energy, are noticeable near the corresponding threshold 
energy for primaries. In this context, it can be important to
remind the considerations expressed
in Sect. 2, about the importance of high $X_F$ region in meson
production, where the distribution from different models depart
more one from another.

A summary concerning the quantities affecting the transverse structure of the high energy
muon component is reported in Table \ref{tb:t3}.
These are the average depth of the first interaction $X_{first}$, $<P_\perp>$
for pions coming from the first interaction, the average slant production height
$H_{\mu}$ of muons survived underground (the decay height of their parent
mesons\footnote{CORSIKA does not allow to access directly to the production
height of parent mesons, which would be more interesting for our purposes}), the
average distance of muons from shower axis underground ($<R>$) and the average
muon pair separation for multiple muon events ($<D>$).

\begin{table} [p]
\catcode`?=\active \def?{\kern\digitwidth}
\caption{Comparison of a few relevant quantities concerning the
lateral distribution of underground muons at the depth of 3400
hg/cm$^2$, from proton primaries at 20, 200 and 2000 TeV, 
30$^\circ$ zenith
angle. 
\label{tb:t3}} 
\begin{tabular*}{\textwidth}{@{}l@{\extracolsep{\fill}}ccccc}
\hline
                  \multicolumn{6}{c}{p--Air, 20 TeV} \\
\hline
Code    & $<X_{first}>$  & $<P_\perp>$ $\pi^\pm$  & $<H_{\mu}>$ & 
$<R>$ & $<D>$ \\ 
                & (g/cm$^2$) & (GeV/c) & (km)  & (m)    &  (m)  \\
\hline
HEMAS           &  51.4   &  0.40  & 24.1 &  7.9  & 12.7 \\
\hline 
CORSIKA/DPMJET  &  44.4   &  0.42  & 25.6 & 10.1  & 13.9 \\
\hline
CORSIKA/QGSJET  &  45.7   &  0.39  & 24.3 &  7.3  & 10.0 \\
\hline
CORSIKA/VENUS   &  48.3   &  0.35  & 24.5 &  7.4  &  8.3 \\
\hline 
CORSIKA/SIBYLL  &  50.9   &  0.37  & 23.5 &  7.2  & 11.5 \\
\hline
\end{tabular*}
\begin{tabular*}{\textwidth}{@{}l@{\extracolsep{\fill}}ccccc}
\hline
                  \multicolumn{6}{c}{p--Air, 200 TeV} \\
\hline
Code  & $<X_{first}>$  & $<P_\perp>$ $\pi^\pm$  & $<H_{\mu}>$ & 
$<R>$ & $<D>$ \\ 
                & (g/cm$^2$) & (GeV/c) & (km)  & (m)    &  (m)  \\
\hline
HEMAS           &  56.1   &  0.44  & 20.6 &  5.3 & 8.0 \\
\hline 
CORSIKA/DPMJET  &  53.9   &  0.43  & 21.7 &  6.2 & 8.8 \\
\hline
CORSIKA/QGSJET  &  52.8   &  0.41  & 21.4 &  5.5 & 7.8 \\
\hline
CORSIKA/VENUS   &  60.2   &  0.36  & 20.9 &  5.3 & 7.5 \\
\hline 
CORSIKA/SIBYLL  &  55.2   &  0.41  & 20.2 &  5.2 & 7.3 \\
\hline
\end{tabular*}
\begin{tabular*}{\textwidth}{@{}l@{\extracolsep{\fill}}ccccc}
\hline
                  \multicolumn{6}{c}{p--Air, 2000 TeV} \\
\hline
Code & $<X_{first}>$  & $<P_\perp>$ $\pi^\pm$  & $<H_{\mu}>$ & 
$<R>$ & $<D>$ \\ 
                & (g/cm$^2$) & (GeV/c) & (km)  & (m)    &  (m)  \\
\hline
HEMAS           &  63.0   &  0.50  & 16.3 &  4.1  & 6.0 \\
\hline 
CORSIKA/DPMJET  &  60.0   &  0.42  & 18.5 &  4.9  & 6.4 \\
\hline
CORSIKA/QGSJET  &  63.1   &  0.44  & 17.7 &  4.2  & 5.6 \\
\hline
CORSIKA/VENUS   &  66.7   &  0.36  & 16.8 &  4.1  & 5.3 \\
\hline 
CORSIKA/SIBYLL  &  60.3   &  0.44  & 17.0 &  4.4  & 5.6 \\
\hline
\end{tabular*}
\end{table}

In the energy range around hundreds of TeV, to which most of data
taken at Gran Sasso laboratory 
belong, the resulting differences in the average muon separation do not exceed
20\%. These discrepancies seem to reduce at higher energy, while they appear
much larger a few tens of TeV.
DPMJET is probably the only model predicting a higher average separation than
HEMAS. 
A precise analysis of the reasons leading to those differences in the models
is complicated, however, it is important to notice that, in fact, HEMAS gives
in general higher values of
average $P_\perp$ with respect to the other models. The only
exception is indeed DPMJET, which, as mentioned before, 
has a particular attention
to the reproduction of nuclear effects affecting the transverse momentum, as
measured in heavy ion experiments. On the other hand, the effect on
this larger $P_\perp$ on the lateral distribution of 
muons is moderated in HEMAS by a
deeper penetration of showers, giving 
in general a somewhat smaller average height of meson production.
Similar features in the comparison of models are also obtained 
for nuclear projectiles.
The data analysed by MACRO\cite{macrodeco1,macrocomp2} favour the HEMAS
simulation, however a comparison of the same data with other models is
still missing.

\section{Conclusions}

It has been shown that, as far as TeV muons are concerned,
differences in the predicted yield are not enormous, although
they exist. So far, the published analyses from different
experiments have been mostly based on a given specific model.
In the light of the present results, it is instead 
recommended that each group should consider more than one model for
the interpretation in terms of spectrum and composition of primaries. 
In particular, it would be highly desirable
that the set of considered simulation codes for event generation
would be the same for every collaboration.
This of course requires a considerable
effort to the experimental groups, however it appears necessary
in view of a sensible comparison of the different results, which
at present are sometimes still controversial, despite the high
quality reached in the detector and analysis techniques.
This is valid not only for the underground
experiments, but also for EAS array.
In this respect, more complete results
on code comparison will be provided in a next work\cite{aabfr}.
The results presented at this conference by the MACRO experiment\cite{ornella},
are an important example in the in the case of DPMJET, which provides a larger muon yield underground.
suggested direction: the analysis of primary spectrum,
originally based on HEMAS, can converge to significantly different values
in the case of DPMJET, which provides a larger muon yield underground.

A fundamental question is if it is possible an experimental
discrimination of hadronic interaction models on the basis of
cosmic ray data themselves. As far as high energy muons
are concerned, the answer is not yet certain. However,
there are promising interesting measurements which
cover the range of primary energy below few tens of TeV.
The coincidence experiment Ne--N$_\mu$
at Gran Sasso\cite{coinc2} has been already mentioned.
The same collaborations 
are now preparing the analysis of coincidences
between Cherenkov signal and TeV muons underground\cite{morello}.
This would allow the study of TeV muon yield as a function of
energy in the threshold region. It is important
to remark that, although these check can be done
in a limited and relatively low energy, they are in any case
important to understand features, like the
shape of $X_{lab}$ distribution which, as discussed before, 
are essentially preserved in a wide energy range.

Last but not least, the quality of the different models can be 
debated just on phenomenological bases. 
Despite the common basic language, the presently available 
``theoretically inspired'' models do still contain many free parameters and
rather different choices for the algorithmic solutions. 
Hopefully, if future data on soft physics from LHC 
will be available (after year 2005), these might be helpful
in reducing some of the uncertainties in the interaction features
and in the shower calculations,
as far as the energy region of the knee, and above, is concerned.

\section*{Acknowledgments}

The author is indebted to 
M. Ambrosio, C. Aramo, M. Carboni, C. Forti
and J. Ranft for the technical help with the 
codes and for the critical discussions.


\begin{thebibliography}{9}
\bibitem{auger}  A.A.~Watson. Nucl. Phys. B (Proc. Suppl.) {\bf 60B} (1998) 171
\bibitem{kaskade} P.~Doll et~al., 
KFK 4686, Karlsruhe report,  1990
\bibitem{corsika} J.N.~Capdevielle et al., Karlsruhe report KFK 4998 (1992).
\bibitem{karls} J.~Knapp, D. Heck and G. Schatz, Karlsruhe report FZKA 5828 (1996).
\bibitem{knapp} J.~Knapp, these proceedings.
\bibitem{utah} H.E.~Bergeson et al., Phys. Rev. Lett. {\bf 35} (1975) 1681.
\bibitem{nusex} G.~Bologna et al., Il Nuovo Cimento {\bf 8C} (1985) 76.
\bibitem{frejus} Ch.~Berger et al., Phys. Rev. {\bf D40} (1989) 2163.
\bibitem{macrocomp} MACRO Collaboration,
Phys. Rev. D {\bf 46} (1992) 895.
\bibitem{macrocomp2} MACRO Collaboration, M.~Ambrosio et al., Phys. Rev.
{\bf D56} (1997) 1407 and 1418.
\bibitem{soudan2} SOUDAN2 Collaboration, Phys. Rev. {\bf D55} (1997) 5282.
\bibitem{kgf} H.R.~Adakar et al., Phys. Rev. {\bf D57} (1998) 2653.
\bibitem{lvd} LVD Collaboration, Il Nuovo Cimento {\bf 105A} (1992) 1815 and
Astropart. Phys. {\bf 2} (1994) 103.
\bibitem{eastop} EAS--TOP Collaboration,  
Nucl. Instr. \& Meth. {\bf A277} (1989) 23.
\bibitem{coinc} EAS-TOP and MACRO Collaborations,  Phys. Lett., {\bf B337
} (1994) 376.
\bibitem{aabfr} C.~Aramo, M.~Ambrosio, G.~Battistoni and C.~Forti, 
work in preparation.
\bibitem{vulc_gais} T.K.~Gaisser, Proc. Vulcano Workshop 1992 on ``Frontier
Objects in Astrophys. and Particle Phys.'', {\bf 40}, (1993) 433. 
\bibitem{muontransp} G.~Battistoni et al., Nucl. Instr. \& Meth. 
{\bf A394} (1997) 136.
\bibitem{coinc2} EAS-TOP and MACRO Collaborations,
Proc. 25th I.C.R.C., Durban, {\bf 6} (1997) 85.
\bibitem{elbert} J.W.~Elbert, Proc. DUMAND Summer Workshop, Vol. {\bf 2} (1978) 101;
J.W.~Elbert et al., Phys. Rev. {\bf D27} (1983) 1448.
\bibitem{nim85} T.K~Gaisser and T.~Stanev, Nucl. Instr. \& Meth. {\bf A235}
(1985) 183.
\bibitem{hillas} A.M.~Hillas, Proc. 17th I.C.R.C., Paris, {\bf 8} (1981) 193.
\bibitem{hemas} C.~Forti et al., Phys. Rev. {\bf D42} (1990) 3668.
\bibitem{macrodeco1} MACRO Collaboration, S.P.~Ahlen et al., Phys. Rev.
{\bf D46} (1992) 4836.
\bibitem{UA5} G.J.~Alner et al., Physics Letters {\bf 167B} (1986) 476.
\bibitem{charm} G.~Battistoni et al., Astropart. Phys. {\bf 4} (1996) 351.
\bibitem{dpm} A.~Capella et al., Phys. Rep. {\bf 236} (1994) 225. 
\bibitem{grt} V.N.~Gribov, Sov. Phys. JETP {\bf 26} (1968) 414; 
V.N.~Gribov and A.A.~Migdal,
Sov. J. Nucl. Phys. {\bf 8} (1969) 583.
\bibitem{glauber} R.J~Glauber and G.~Matthiae, Nucl. Phys. {\bf B21} 135.
\bibitem{lund} H.~Bengtsson and T.~Sj\"{o}strand,
Comput. Phys. Commun. 46 (1987) 43.
\bibitem{gheisha} H.~Fesefeldt, Aachen preprint PITHA 85/02 (1985).
\bibitem{sibyll} R.S.~Fletcher et al., Phys. Rev. {\bf D50} (1994) 5710.
\bibitem{dpmjet} J.~Ranft, Phys. Rev. {\bf D51} (1995) 64. 
\bibitem{ornella} O.~Palamara for the MACRO collaboration, these proceedings.
\bibitem{hemasdpm} G.~Battistoni et al., Astropart. Phys. {\bf 3} (1995) 157.
\bibitem{qgsjet} N.N.~Kalmikov et al, 
Physics of Atomic Nuclei {\bf 58} (1995) 1728.
\bibitem{qgs} A.B.~Kaidalov et al., Yad. Fiz. 43 (1986) 1282.
\bibitem{venus} K.~Werner, Phys. Rep. {\bf 232} (1993) 87.
\bibitem{stanev} T.~Stanev, these proceedings.
\bibitem{popcorn} T.~Sj\"{o}strand, CERN-TH.6488/92 (1992)
\bibitem{zmom} R.S.~Fletcher et al., Proc. 23rd I.C.R.C., Calgary, {\bf 4} (1993) 40.
\bibitem{gaisser97} G.M.~Frichter, T.K.~Gaisser and T.~Stanev, astro-ph/9704061
and Phys.Rev. {\bf D56} (1997) 3135.
\bibitem{lipstan} P.~Lipari and T.~Stanev, Phys. Rev. {\bf D44} (1991) 3543.
\bibitem{morello} C.~Morello for the EAS--TOP collaboration, these proceedings.
\end{thebibliography}
\end{document}